\documentclass[12pt]{article}
 \usepackage{graphicx}
 \usepackage[cp1251]{inputenc} %% Windows
 \usepackage[russian]{babel}   %%

\begin{document}
 \noindent {\footnotesize\it Astrophysics, Vol. 57, No. 4, pp. 583-604, December, 2014.
  %%\\  Translated from Astrofizika, Vol. 57, No. 4, pp. 625-650 (November 2014).
  }

 \newcommand{\dif}{\textrm{d}}
 \noindent
 \begin{tabular}{llllllllllllllllllllllllllllllllllllllllllllll}
 & & & & & & & & & & & & & & & & & & & & & & & & & & & & & & & & & & & & & \\\hline\hline
 \end{tabular}

 \bigskip
 \leftline{REVIEWS}
 \bigskip
 \centerline{\bf THE GOULD BELT}
 \bigskip
 \centerline{\bf V.V. Bobylev$^{1,2}$}
 \bigskip
 \centerline{\small \it $^1$Pulkovo Astronomical Observatory, St. Petersburg,  Russia}
 \centerline{\small \it $^2$Sobolev Astronomical Institute, St. Petersburg State University, Russia}
 \bigskip
 \bigskip
{\bf Abstract}—This review is devoted to studies of the Gould belt
and the Local system. Since the Gould belt is the giant
stellar-gas complex closest to the sun, its stellar component is
characterized, along with the stellar associations and diffuse
clusters, cold atomic and molecular gas, high-temperature coronal
gas, and dust contained in it. Questions relating to the kinematic
features of the Gould belt are discussed and the most interesting
scenarios for its origin and evolution are examined.

 %DOI 10.1007/s10511-014-9360-7

\section{Historical information}
Stars of spectral classes O and B that are visible to the naked
eye define two large circles in the celestial sphere. One of them
passes near the plane of the Milky Way, while the second is
slightly inclined to it and is known as the Gould belt. The
minimum galactic latitude of the Gould belt is in the region of
the constellation Orion, and the maximum, in the region of
Scorpio-Centaurus.

Herschel noted [1] that some of the bright stars in the southern
sky appear to be part of a separate structure from the Milky Way
with an inclination to the galactic equator of about 20$^\circ$.
Commenting on the features of the distribution of stars in the
Milky Way, Struve [2] independently noted that the stars that form
the largest densifications on the celestial sphere can lie in two
planes with a mutual inclination of about 10$^\circ$.

Gould [3,4] made a detailed study of this structure, determined
the galactic coordinates of the pole of a large circle of the
celestial sphere along which the stars are in groups, and
determined the coordinates of the nodes. He found the inclination
of the major plane of stars to 4$^m$ with respect to the galactic
plane to be 17$^\circ$. Because of his work, the belt is referred
to as the ``Gould belt''.

Shapley [5] provided a clear formulation of the fact that, as
opposed to the distant spherical clusters which delineate the
Galaxy, the close (to 1 kpc) bright stars form a unique separate
subsystem of the same type as a local cluster or local system. He
proposed that this densification of near, bright B stars be called
the Local system.

From a modern standpoint, the term ``Local system'' is more
substantive, since it incorporates not only individual nearby
stars of different spectral classes, but also the system of nearby
stellar associations and diffuse nebulae, the presence of cold
atomic HI, molecular H$_2,$ and high temperature coronal gas and
dust.

A number of kinematic features of the bright O and B stars were
clarified when the first data on the radial velocities, proper
motions, and distances of the stars became available. First, a
constant term of magnitude about $+5$~km/s was discovered. This
feature was first noticed by Frost and Adams in 1903 and confirmed
by Kapeteyn and Frost [6] in an analysis of the radial velocities
of the ``Orion'' stars. Following the suggestion of Campbell [7],
this term is known as the K-term or the K-effect. The presence of
the positive K-term is mostly interpreted as a common expansion of
the stellar system [8,9] and only a small fraction of it
($\approx$1.5~km/s) can be explained as a red shift of spectrum
lines owing to the gravitation of massive stars in accordance with
the theory of relativity. Second, it was shown that the residual
velocities of the O and B-stars typically have a small dispersion
in their residual velocities of magnitude $\leq$10~km/s, as well
as a significant deviation by 20--30$^\circ$ of the vertex from
the direction to the center of the Galaxy [10,11,12].

The papers of Lindblad [13,14] and Oort [15,16] have had a
significant influence on studies of the structure and kinematics
of the Galaxy. The existence of interstellar absorption of light
was definitively established [17] and many authors contributed to
developing models of stellar evolution; O and T associations were
discovered [18,19].

This has ultimately led to an understanding that the stars
belonging to the Gould belt are not only nearby, but also very
young objects (younger than $\approx$60 million years) that are
participating in the differential rotation of the Galaxy, and
their galactic orbits are close to circular. It has been shown in
a number of papers that the observed residual velocities of the
stars in the Gould belt can be interpreted either as a residual
rotation of the entire system or as a combination of rotation and
expansion.

The early history of research on the Gould belt is reflected in
the book by Bok [20]. A systematic discussion of the key questions
relating to studies of the Gould belt, along with a detailed
bibliography, can be found in the work of Frogel and Stothers
[21], Efremov [22], and P\"{o}ppel [23].

\section{Structure}
The Gould belt is the closest giant stellar-gas complex to the
sun. Complexes of this sort are regions of active star formation
and are not only observed in our Galaxy [24], but also in other
galaxies [22,25].

Based on modern estimates, the Gould belt is a quite flat system
with semiaxes of $\approx350\times250\times50$ pc and is inclined
to the galactic plane by 16$^\circ$--22$^\circ$. The ascending
node is in the $l_\Omega=275^\circ$--295$^\circ$ direction. The
sun lies at a distance of about 40 pc from the line of nodes. The
center of the system is located at a distance of about 150 pc in
the second galactic quadrant. More precise knowledge of the
direction to the center l0 depends on the age of the sample, and
ranges from 130$^\circ$ to 180$^\circ$ according to various
estimates.

The spatial distribution of the stars is highly nonuniform; a
significant drop in their density is observed at a radius of
$\approx$80 pc from the center, i.e., the entire system has a
donut shape. The well-known diffuse cluster $\alpha$Per (Melotte
20) with an age of about 35 million years lies near the center of
this donut.

\subsection{Stellar composition}
P\"{o}ppel has pointed out [26] that only the relatively near
stars no later than spectral class B2.5 can be said reliably to
belong to the Gould belt. There are, however, very few of them. In
order to assign stars from other spectral classes to the belt,
various methods of distinguishing the stars from different
populations in the vicinity of the sun are required. In recent
decades x-ray data from the
 ROSAT \footnote {ROentgen SATellite, operated from 1990-1999.},
 Chandra \footnote {X-ray telescope, operating in orbit since 1998.}, and
 XMM-Newton \footnote {X-ray telescope, operating in orbit since 1999.}
satellites, earthbound infrared photometric data from 2MASS [27],
and the HIPPARCOS [28] and Tycho-2 [29] astronomical catalogs have
been used.

Torra, et al. [30], have analyzed the extensive HIPPARCOS sample
of stars in spectral classes O and B that lie within a radius of
$r<2.0$ kpc of the sun. The individual age of 2864 stars was
estimated by Str\"{o}mgren photometry. The relative error in the
age determination for 88\% of the stars in the sample was less
than 100\%. It was concluded that within a radius of $r\leq0.6$
kpc, about 60\% of the stars younger than 60 million years belong
to the Gould belt.

Chereul, et al. [31], have estimated the age of 1077 near
($r<0.125$ kpc) HIPPARCOS stars of spectral classes A and F based
on Str\"{o}mgren photometry. The average error in the age
determination was about 30\%, but for the youngest fraction it
reached 100\%. The distribution of AF stars with respect to age
has two peaks: the overwhelming majority of the stars are
concentrated in the neighborhood of a peak at 650 million years
and about 400 stars, in the neighborhood of a peak at 10 million
years. We may conclude that 20--30\% of the near AF stars may
belong to the Gould belt. This estimate agrees with the earlier
result of Taylor, et al. [32].

Several hundred T Tau stars have been found in the nearby
associations and diffuse nebulae of the Gould belt. These are
dwarfs in late spectral classes with masses of $\approx1M_\odot$
and ages of several million years that have not reached the Main
sequence stage. Stars have also been identified in terms of their
emission in the HЈ\ line, their lithium abundance, x-ray emission,
position in the Hertzsprung-Russell diagram, rapid axial rotation
(v~sin\,i$\approx$30 km/s an higher), and kinematic
characteristics.

X-ray emission in the 0.18--0.3 keV range (ROSAT) makes it
possible to detect sources with temperatures below $10^6$~K
(approximately later than G0 for Main sequence stars). A
significant lithium abundance indicates that an already formed
star is in a short-duration state in which thermonuclear reactions
have not yet begun. Thus, when there is more lithium, a star is
younger. A comparison of the position of these stars on a
Hertzsprung-Russell diagram with appropriate isochrones makes it
possible to select the youngest objects. A relationship between
the lithium abundance, age, and kinematics of the closest young
dwarfs has been demonstrated by Wichman, et al. [33].

Figure 1 is an illustrative sample of candidate T Tau stars from
the Scorpio-Centaurus association based on data from Ref.~[34]. In
the left frame the candidates are indicated by open circles and
crosses, and in the right, by solid circles.

The T Tau stars are divided into several categories: classical
(CTTS, classical T Tauri stars) are the youngest objects with ages
less than 10 million years and they still have a dust disk; stars
with less marked characteristics (WTTS, weak-line T Tauri stars)
are older and lie closer to the Main sequence; and the oldest
stars of this type (PTTS, post-T Tauri stars).

There is also some interest in the very young massive
star-formation regions with an emission spectrum--Herbig--Haro
objects (a wide class of stars designated as HAeBe) which are also
in a stage of not having reached the Main sequence. About ten of
these have been discovered among the nearby OB associations
related to the Gould belt [35].

We note that about 40 unidentified g-ray sources with energies
above 100 MeV have been detected (using the EGRET system on board
CGRO\footnote {Compton Gamma-ray Observatory, a satellite that
operated during 1991--2000.}). Their distribution has a
statistical relation to the Gould belt, but the nature of these
sources is still not clear.

%%%%%%%%%%%%%%%%%%%%%%%%%%%%%%%%%%%%%%%% f.1:
 \begin{figure}[t]
 {\begin{center}
 \includegraphics[width=120mm]{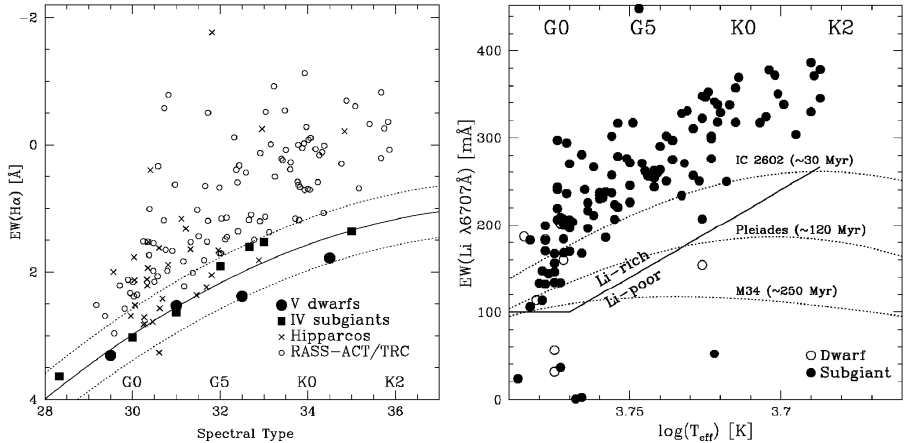}
 \caption{
 An example of samples of candidate T Tau stars from the Scorpio-Centaurus
association based on equivalent width (EW) data for the H$_\alpha$
emission line as a function of spectral class (left) and for the
Li I $\lambda$6707\AA line as a function of effective temperature
(right) according to the data of Mamajek, et al. [34].}
 \label{fig1}
 \end{center} }
 \end{figure}
%%%%%%%%%%%%%%%%%%%%%%%%%%%%%%%%%%%%%%%%%%%%%%%

\subsubsection{Moving groups and diffuse stellar clusters}
Diffuse stellar clusters lie within a wide region near the sun.
Their distances and ages have been estimated reliably, so they are
of great interest for studying the kinematics of the Galaxy, in
particular of the Gould belt. The HIPPARCOS and Tycho-2 catalogs
have made it possible to determine the average values of the
proper motions of the diffuse stellar clusters with fair accuracy.
These data, together with the radial velocities, can be used in a
three-dimensional kinematic analysis.

The COCD catalog [36], which is complete out to $r\approx0.8$ kpc,
is of interest. This catalog has been used to separate diffuse
stellar clusters belonging to the Gould belt from the common
background. Diffuse stellar clusters with ages below 80 million
years were used. This ultimately yielded a sample of 23 diffuse
stellar clusters lying within a radius of $r<0.5$ kpc, for which
the probability of their belonging to the Gould belt was estimated
to be $P_t = 68\%.$ $P_t$ was determined from the kinematics of
the diffuse stellar clusters.

There are definite signs that the recently discovered near and
very young diffuse stellar clusters and moving groups $\beta$~Pic,
TWA, Tuc/Hor, $\eta$ Cha, and $\varepsilon$ Cha also belong to the
Gould belt as members of the Scorpio-Centaurus association or of
its diffuse corona. Assuming that the stars in a diffuse stellar
cluster were formed simultaneously, it is possible to determine
the age of the diffuse stellar cluster by comparing the position
of the stars on a Hertzsprung-Russell diagram with suitable
isochrones.

It has been shown [37,38] that the well-known group parallax
method can be used to determine the individual distances of stars
in nearby (no more than 150 pc from the sun) diffuse stellar
clusters, such as the Hyades, more accurately than from the
HIPPARCOS parallaxes. This, in turn, makes it possible to
``improve'' the Hertzsprung-Russell diagram. According to modern
concepts, as compact gravitationally coupled systems, young
diffuse stellar clusters form part of structures of a larger
spatial scale, i.e., associations.

\subsubsection{Stellar associations}
With the appearance of the first data on the spectral classes of
bright stars, many researchers isolated certain distinct groups of
stars in classes O and B. Later, Ambartsumyan [18] suggested that
these should be called associations.

The hypothesis of a low spatial density and, thereby, the dynamic
instability of associations in the tidal force field of the Galaxy
was also first advanced by Ambartsumyan [19]. He estimated that an
association should disperse over no more than $10^8-10^9$ years.
Blaauw showed [39] that the differential rotation of the Galaxy
causes an association with an initially spherical shape to stretch
out into an ellipse with a time-varying orientation.

The Scorpio-Centaurus association is the closest to the sun.
Blaauw made [40] the first estimate of the kinematic age of this
association, $\approx$20 million years, by analyzing the radial
velocities of the stars using the expansion coefficient $K=50$
km/s/kpc which he found. This age corresponds to the time over
which the star covers the characteristic radius of the region
occupied by the association. A critical review of models for the
formation of the Scorpio-Centaurus association has been written by
Sartori, et al. [41]. Associations are young systems with ongoing
star formation. The closest associations are of undoubted interest
for studying the Gould belt.

A detailed description of the known OB associations within a
radius of $\approx$1.5 kpc, including ones belonging to the
structure of the Gould belt, by Zeeuw, et al. [42], makes use of
the probable members among selected HIPPARCOS stars. This list
included the following associations: Cep OB2, Lac OB1, Cep OB6,
Per OB2, Cas-Tau, Sco-Cen (US, UCL, LCC), Tr 10, Vel OB2, and Col
121. A description of and the stellar composition of the
association Ori OB1 can be found in the paper by Brown, et al.
[43]. The distribution of these associations in the galactic plane
is shown in Fig. 2.

%%%%%%%%%%%%%%%%%%%%%%%%%%%%%%%%%%%%%%%% f.2:
 \begin{figure}[t]
 {\begin{center}
 \includegraphics[width=100mm]{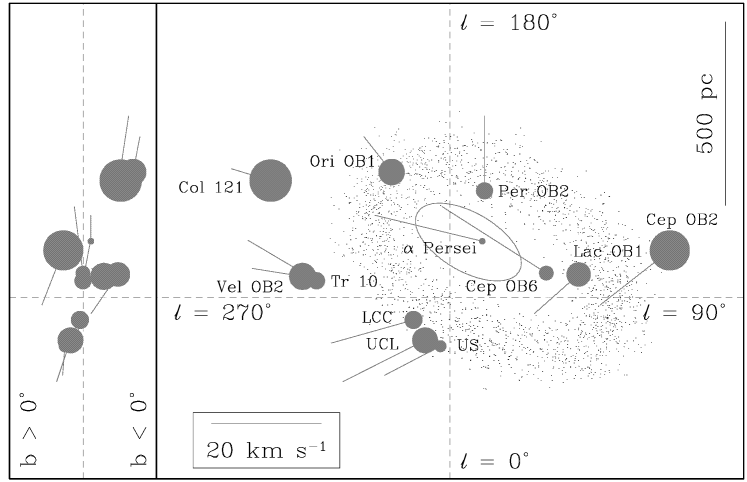}
 \caption{
 The spatial distribution of nearby OB associations projected onto the
galactic plane. The circles denote the sizes of the associations.
The residual velocity vectors relative to the local standard of
rest, free from the differential rotation of the Galaxy, are
indicated. This figure is taken from Ref. 42. The ellipse around
the diffuse stellar cluster $\alpha$Per indicates the association
Cas-Tau. The cloud of points is a schematic representation of
Olano’s model [44] of the Gould belt.}
 \label{fig2}
 \end{center} }
 \end{figure}
%%%%%%%%%%%%%%%%%%%%%%%%%%%%%%%%%%%%%%%%%%%%%%%

It has been shown [45,22] that the OB associations and young
clusters (B2 and younger) within 3 kpc of the sun may be combined
into complexes with sizes of 150--700 kpc. Almost all of them
contain giant molecular clouds with masses $\geq10^5 M_\odot$.
Many of the complexes are coupled to giant clouds of neutral
hydrogen. The Gould belt is one of these giant complexes.

 \subsubsection{The Local system of stars, Supercluster, Supercomplex}
 According to Mineur [46,47], besides participating
in the common galactic rotation, the Local system of stars gives
signs of rotating around a center that does not coincide with the
center of the Galaxy. This idea was examined with more extensive
data by Shatsova [48]. She found an intrinsic rotation of the
Local system that was in the same direction as the galactic
rotation. Here the Local system was treated as the set of all
nearby stars. The problem of separating the $\approx$33000 stars
of mixed spectral composition in the Boss catalog into fractions
(according to signs of participation in the galactic rotation) was
not addressed. Thus, as noted by Shatsova, her results were of a
preliminary nature. The size of the region of space being analyzed
and specified by the Boss catalog was estimated to be 300--350
kpc. The kinematics and dynamics of the Local system occupy a
special place in the book by Ogorodnikov [9].

Tsvetkov [49,50,51] undertook further study of the Local stellar
system of stars based on Shatsova's equations. He showed, first of
all, that the systematic errors in the GC, N30, FK4, and FK5
catalogs did not lead to significant differences in the parameters
of the Local system. Second, with data from the HIPPARCOS catalog
he obtained solutions separately for groups of stars subdivided
according to spectral class and to distance from the sun.
Ultimately, it was possible to localize the Local system on a
Hertzsprung-Russell diagram as a system of stars of spectral
classes A-F in the Main sequence, formed near the center, and
lying in the $l=253^\circ\div9^\circ$, $b=-13^\circ\div9^\circ$
direction at a distance of 180 pc from the sun. The rotation is
counter-clockwise, i.e., in a direction opposite to the galactic
rotation, with a period of about 140 million years in a plane
inclined to the plane of the galaxy by 30ўX. The Local stellar
system has no effect on the motion of stars at distances greater
than 300 pc from the sun.

The idea of the Supercluster was developed by Eggen [52,53], who
combines several fairly young clusters into the Pleiades group
(age 100--150 million years) based on the closeness of their
kinematics and names this group the ``Local association''. The
spatial velocities of the stars (calculation of which required
high precision parallaxes) were used to select the members of the
association. One gets a sense that the group was not selected
entirely correctly. In fact, in those years high precision
trigonometric parallaxes existed only for a very small
($\approx$50 pc) neighborhood of the sun, so only the nearest
stars were used. Eggen's lists of the members of the Pleiades
group (the current name for the Local association) are used, for
example, in Refs. 54 and 55.

Barkhatova, et al. [56], analyzed the properties of a number of
isolated diffuse stellar clusters of different ages near the sun
and advanced the hypothesis that they belong to a higher order
system, the Supercomplex. The diameter of the Supercomplex was
2000 pc, its thickness was 150 pc, and 11 complexes of diffuse
stellar clusters and 4 isolated diffuse stellar clusters were
assigned to it. It was shown that the Supercomplex has a residual
rotation with an angular velocity of $\approx$12 km/s/kpc in the
same direction as the galactic rotation. The Supercomplex has a
size comparable to the Orion arm. The idea that the Gould belt is
part of the Orion arm was discussed some time ago, but its most
complete expression is to be found in the dynamic model of Olano
[44], which will be discussed in more detail in section 3.3.

 \subsection{Interstellar medium}
 \subsubsection{Neutral hydrogen HI}
 An analysis of the earliest
observations of neutral hydrogen at 21~cm showed that its
distribution is related to the Gould belt [57,58].

Lindblad, et al. [59,60] have shown that the motion of nearby
hydrogen clouds has the same kinematic features as for the Gould
belt --- a common expansion effect. This has been confirmed by
others. Lindblad isolated and studied a region which he referred
to as ``detail~A,'' which is still known as the Lindblad ring. The
spatial dimensions of the ring are $\approx800\times500$ pc [61]
and its center lies in the second galactic quadrant, close to the
presumed center of the Gould belt. This implies that the Gould
belt is surrounded by a giant cloud of neutral hydrogen.

%%%%%%%%%%%%%%%%%%%%%%%%%%%%%%%%%%%%%%%% f.3:
 \begin{figure}[t]
 {\begin{center}
 \includegraphics[width=90mm]{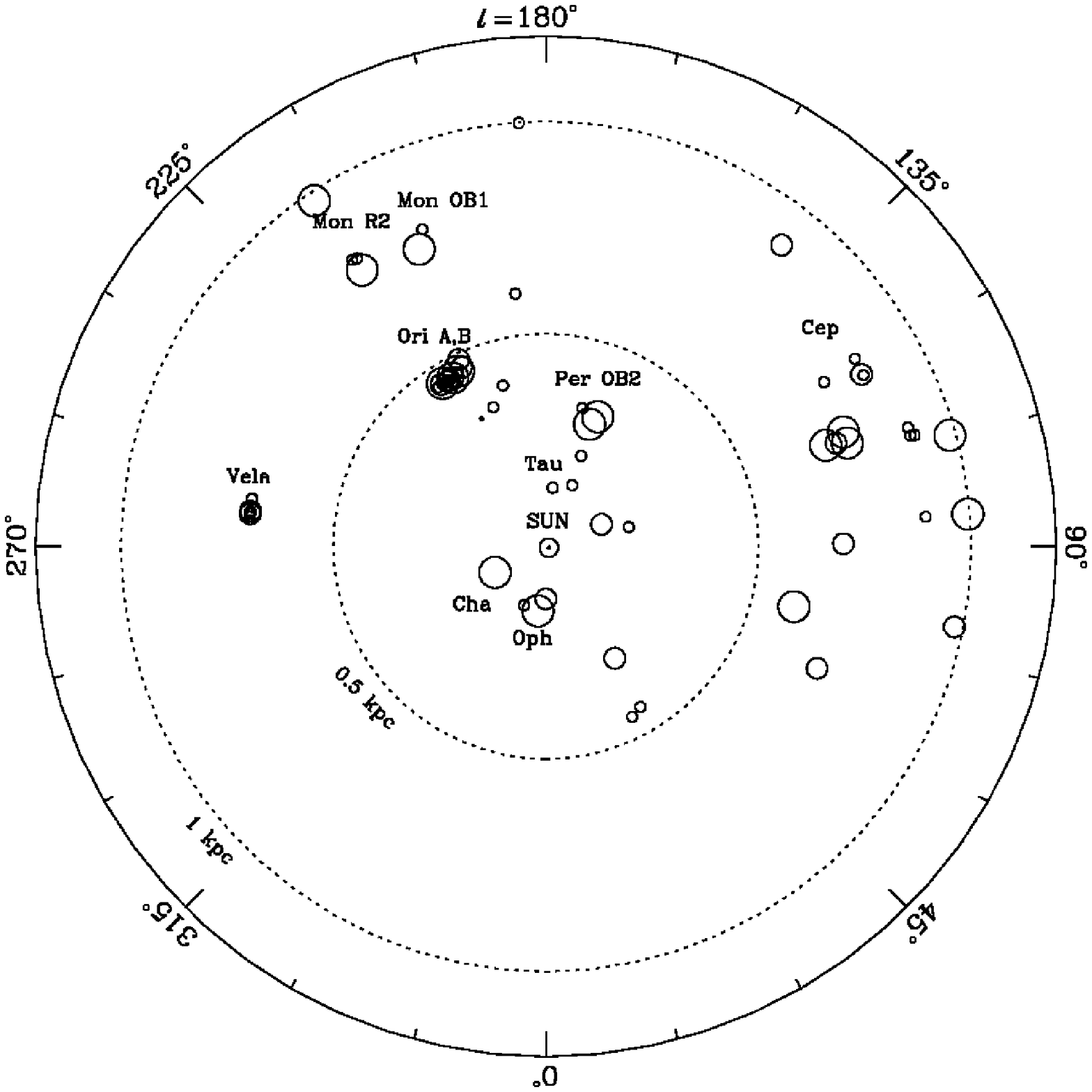}
 \caption{
 The spatial distribution of young stellar groups and diffuse stellar
clusters within a radius of 1 kpc from the sun according to the
data of Porras, et al. [64], with the large complexes of molecular
clouds indicated.}
 \label{fig3}
 \end{center} }
 \end{figure}
%%%%%%%%%%%%%%%%%%%%%%%%%%%%%%%%%%%%%%%%%%%%%%%
%%%%%%%%%%%%%%%%%%%%%%%%%%%%%%%%%%%%%%%% f.4:
 \begin{figure}[t]
 {\begin{center}
 \includegraphics[width=80mm]{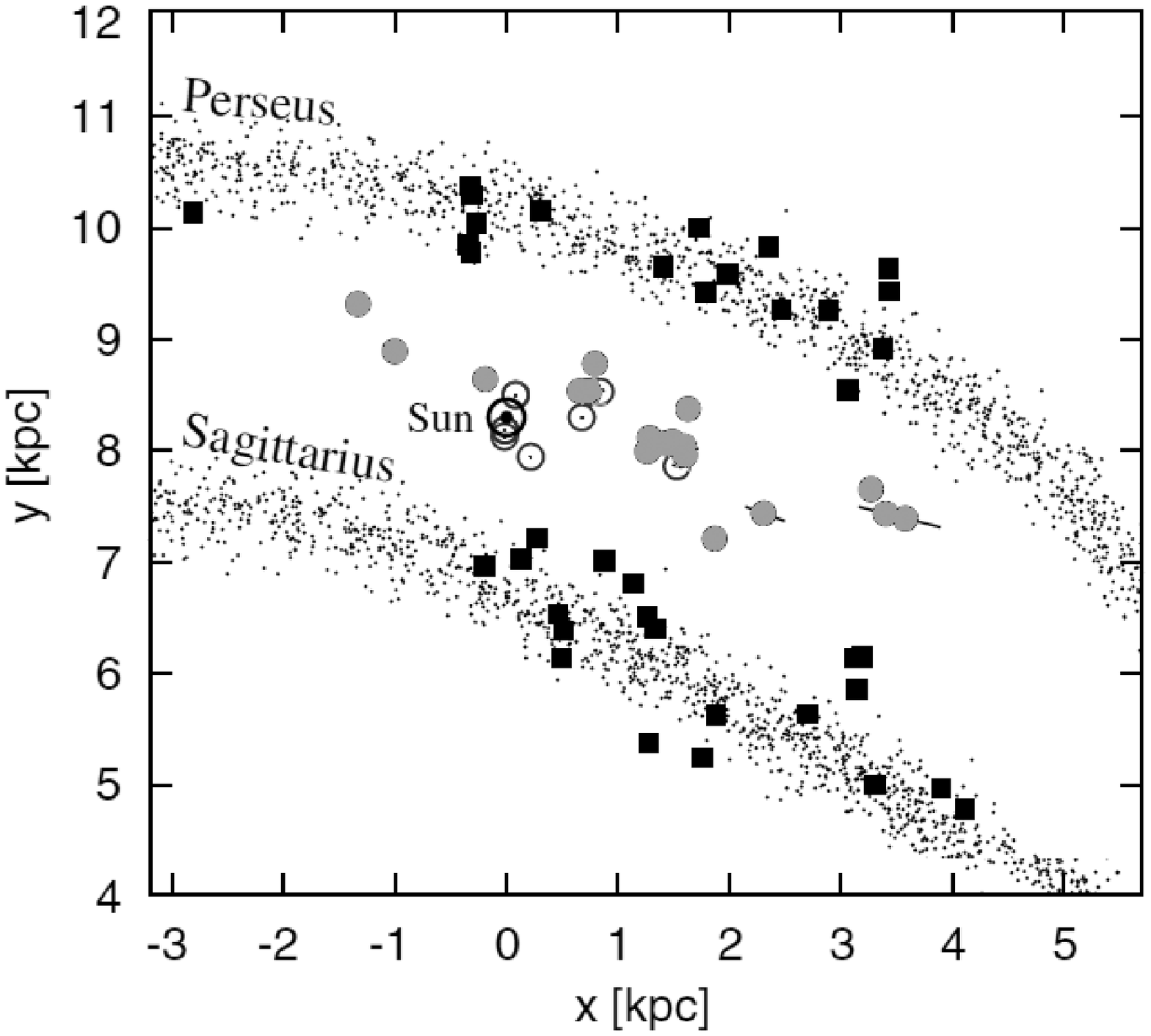}
 \caption{
 Masers with measured trigonometric parallaxes
in segments of the spiral arms of Perseus and Carina- Sagittarius
(squares) nearest the sun and in the local arm (circles) according
to the data of Xu, et al. [72].}
 \label{fig4}
 \end{center} }
 \end{figure}
%%%%%%%%%%%%%%%%%%%%%%%%%%%%%%%%%%%%%%%%%%%%%%%

\subsubsection{H2 molecular clouds and HII zones}
Direct study of the distribution of molecular hydrogen in the
Galaxy is difficult because it has no emission lines in the radio
range. There is a reliable indirect method based on the relative
content of carbon monoxide CO and molecular H2, which equals
$6\times10^{-5}.$ Thus, the 2.6~mm CO radio line turned out to be
a convenient indicator of the distribution of molecular hydrogen.
A detailed discussion of the methods for observing the CO line and
a catalog of molecular clouds can be found in a paper by Dame, et
al. [62].

The relation between the Gould belt and low-latitude
($|b|\leq24^\circ$) molecular clouds and their distributions and
kinematics have been studied by Taylor, et al. [32], who show that
the motion of the clouds is consistent with a model of expansion
of the Gould belt. An analysis of the distribution over the
celestial sphere of high-latitude ($|b|\geq25^\circ$) molecular
clouds [63] shows that they form two extended shells associated
with the two closest OB associations in the Gould belt: Per
OB3/Cas-Tau and Sco-Cen.

There is a very close relationship between molecular clouds and
star-formation regions in the Galaxy. This is because stars are
formed inside these clouds. Zones of ionized hydrogen develop
around very young and massive O or B stars and these are
indicators of star-formation regions. One of the best known HII
zones, the Orion nebula (the association Ori OB1), lies in the
Gould belt.

A paper by Porras, et al. [64], provides the most complete
discussion of observational data in the near infrared (J,H,K) of
the youngest stellar groups (with ages of a few million years)
within a radius of 1 kpc from the sun and demonstrates their close
relationship to the distribution of molecular clouds. That
distribution, constructed on the basis of about 7200 stars, is
shown in Fig. 3, where the sizes of the circles correspond to
three categories of groups depending on the number of stars,
 $n<30,$ $30<n<100,$ and $n>100,$ respectively. Complexes of
molecular gases lying within a circle of 0.5~kpc, Ori, Per OB2,
Tau, Cha, Oph, as well as a number of smaller objects, belong to
the Gould belt.

During their formation, protostars have extended shells within
which maser emission is produced. Their trigonometric parallaxes
and proper motions have been determined by very-long base-line
radio interferometry (VLBI) to a very high accuracy, averaging
5--10\%.

Methanol (CH$_3$OH, 6.7 GHz, 12.2 GHz) and water (H$_2$O, 22.2
GHz) masers have been observed on VLBA in the USA [65]. Similar
observations have been made using the European very-long base-line
radio interferometer [66], which includes three Russian antennas:
Svetloe, Zelenchukskaya, and Badary. At present, these two
programs are combined in the BeSSeL \footnote
{http://www3.mpifr-bonn.mpg.de/staff/abrunthaler/BeSSeL/index.shtml}
(Bar and Spiral Structure Legacy Survey, Brunthaler, et al. [67]).
Radio observations of water masers at 22.2 GHz [68] are being
carried out in the Japanese program VERA (VLBI Exploration of
Radio Astronomy). The trigonometric parallaxes of a number of
low-mass nearby stars associated with the Gould belt have been
determined as part of a separate program for the VLBI observations
of radio stars in the continuum [69,70].

At present, the trigonometric parallaxes of more than 100 maser
galactic sources have been measured by various groups of radio
astronomers [71]. These sources are associated with massive active
star formation regions. The observation program includes more than
400 objects. For now, the objects are in the northern hemisphere,
but soon masers will be observed in the southern sky.

Figure 4 shows the distribution of the masers with measured
trigonometric parallaxes in an extensive neighborhood of the sun
within a radius of about 4 kpc. The local arm (the Orion arm),
which traces massive (solid circles) and low-mass protostars (open
circles) can be seen clearly.

\subsubsection{Coronal gas}
An interstellar rarefied hot gas structure with a temperature of
$\approx10^6$~K and a radius of 200--300~pc in the immediate
vicinity of the sun is closely coupled to the Gould belt. It
includes such regions as the ``Local bubble'' and North polar spur
(or Loop I superbubble).

%%%%%%%%%%%%%%%%%%%%%%%%%%%%%%%%%%%%%%%% f.5:
 \begin{figure}[t]
 {\begin{center}
 \includegraphics[width=110mm]{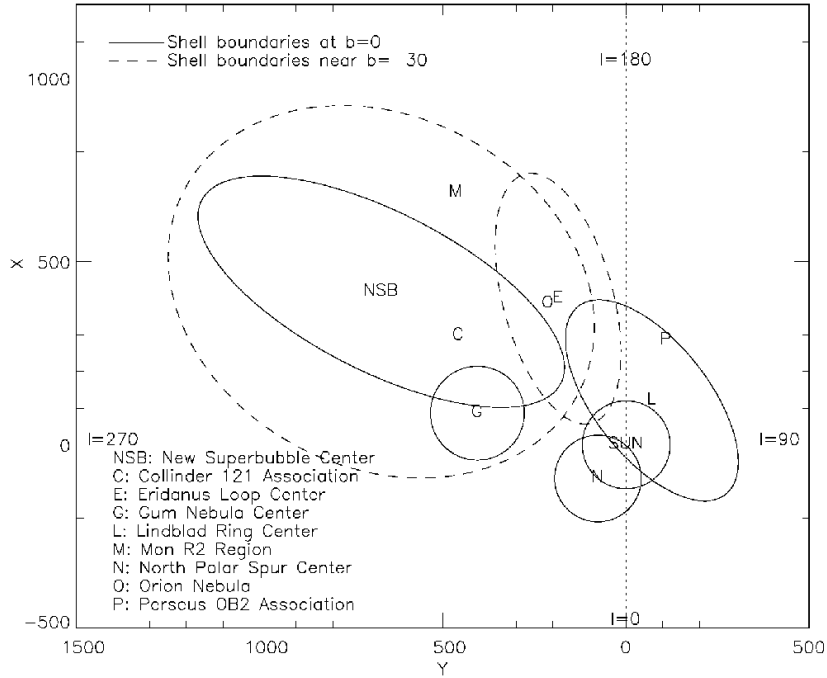}
 \caption{
 Spatial distribution of the regions of coronal gas according to Heiles [82].
The circle with the sun in the center indicates the Local bubble.}
 \label{fig5}
 \end{center} }
 \end{figure}
%%%%%%%%%%%%%%%%%%%%%%%%%%%%%%%%%%%%%%%%%%%%%%%

The local bubble is a compact region that is essentially free of
absorbing material, so it was first discovered in an analysis of
interstellar stellar reddening by Fitzgerald [73]. Charts of the
distribution of neutral gas absorption in NaI lines, constructed
by Sfeir, et al. [74], indicate an inclination to the Galactic
plane, as for the Gould belt.

The physical processes leading to the appearance of bubbles are
generally as follows. In young diffuse stellar clusters there are
many supernova explosions. This leads to the appearance of stellar
winds and the formation of shock waves which sweep gas to the
periphery of a given local region, where it creates densifications
in the form of shells or extended walls. At the boundary of a
shell the gas density increases substantially, the gas cools, and
molecular clouds form. If a bubble is subject to irradiation from
a supernova, the gas in the bubble will be heated and emit x-rays.

Clumps of gas-dust protostellar matter, from which stars will
later be formed, develop in cold fragments of molecular clouds. A
model for subsequent star formation in molecular clouds connected
with OB associations is described by Preibish and Zinnecker [75]
with the Scorpio-Centaurus association as an example.

Berghofer and Breitschwerdt [76] believe that the most realistic
theory for the origin of the Local bubble involves a repeated, not
simultaneous but spread out in time, explosion of about 20
supernovae over the last 10--20 million years. At present, 7
neutron stars are observed in the region of the Gould belt [77,78]
and they could definitely be the residues of such supernovae.

The lifetime of bubbles or caverns are not very long compared to
the age of the Gould belt; this supports the view [79--81] that
the formation of the Local bubble and the North polar loop are
most likely caused by supernova explosions in the
Scorpio-Centaurus association.

Regions of rarefied hot gas of this sort are known within a radius
of about 0.8 kpc of the sun: a nebula in Vela (Gum Nebula), a
complex in Orion-Eridan (associated with the Barnard loop), and
the giant supershell GSH 238+00+09 (NSB new star bubble) recently
identified by Heiles [82]. Their spatial distribution is
illustrated in Fig. 5 where the center of the Lindblad ring, the
OB association Col 121 and Per OB2, and the star formation region
in Monoceros (Mon R2) are indicated (and the Olano model is
plotted). Two contours are shown for the supershell NSB: the
smooth curve indicates a 550$\times$217 pc ellipse at $b=0^\circ$
and the dashed curve, a 605$\times$480 pc ellipse at
$b=-30^\circ$. As Heiles has pointed out, because of errors in the
distance determinations the actual boundaries of the supershell
may be entirely different, especially as regards its extended
profile indicated by the smooth curve in the figure. The origin of
the supershell GSH~$238+00+09$ is still unclear.

A large-scale rarefied hot gas structure called the Great Rift is
known to exist near the sun. Dense clouds of gas and dust lie
along the boundaries of the Great Rift.

\subsubsection{Dust}
The total mass of interstellar dust in the vicinity of the sun is
about 1\% of the overall mass of available hydrogen [57,58]. Dark
dust clouds are a serious problem when estimating the photometric
distances to stars or diffuse stellar clusters. A number of
authors have suggested that the Local system is purely virtual and
shows up because of nonuniformities in the distribution of
absorbing material [36].

Observations of dust concentrated in the disks surrounding
individual young stars are currently of great interest for
recovering the history of star-formation. The JCMT (James Clerk
Maxwell Telescope) project is aimed at solving this problem; it is
to make submillimeter wavelength observations of star-formation
regions belonging to the Gould belt during 2007--2009 [83]. Frisch
[84] has written a detailed review of the properties of the
interstellar medium in the vicinity of the sun.

 \subsection{Kinematics}
 Here we note some results obtained before
and after the HIPPARCOS experiment. Based on a linear
Ogorodnikov-Milne model, Westin [85] has analyzed $\approx$1500
stars in spectral classes O-A0 in the sun's vicinity, as well as
another 500 bright stars. The age of the individual stars was
estimated using 4 color and H$\beta$ Str\"{o}mgren photometry. The
available radial velocities, as well as the proper motions of the
stars from the FK4 catalog, were invoked. The criterion for
belonging to the Gould belt was an age limit for the stars of
$\tau<30$ million years.

Lindblad, et al. [86], have made a similar analysis using
$\approx$2440 HIPPARCOS OB stars, for which Str\"{o}mgren
photometry data are available. Besides the $\tau<30$ million years
criterion, stars lying inside the Lindblad ring were considered to
belong to the Gould belt.

Comeron [87] has used an epicycle approximation method for
analyzing the motions of $\approx$300 of the youngest HIPPARCOS OB
stars along the z axis. The spatial orientation of the axis of the
vertical oscillations was determined in a plane perpendicular to
the galactic plane and found to be 6.5$\pm$1.8 km/s/kpc.

Based on a linear Ogorodnikov-Milne model, Torra, et al. [30],
have made an extensive analysis of $\approx$2500 HIPPARCOS OB
stars drawing on the available data on the radial velocities of
the stars. It was shown that the Oort parameter depends
significantly on the age of the stars and the K-effect was
carefully studied. For OB stars younger than 60 million years, the
expansion velocity reaches 7.1$\pm$1.4 km/s/kpc for an average
sample radius of 100 pc; at larger distances from the sun this
velocity becomes negative with large measurement errors. A special
experiment showed that the expansion effect remains even if the
stars in the Scorpio-Centaurus and Ori OB1 associations are
excluded. This is interesting because OB associations have their
own noticeable intrinsic expansion; the kinematic method for
estimating their age is based on this expansion [40].

%%%%%%%%%%%%%%%%%%%%%%%%%%%%%%%%%%%%%%%%%%%%%%%%%%%%%%%%%%%%%%%%%%%%%%%%%%%%%
{\begin{table}[t]
\small                                                %% t1
\caption[]{\small
 The Oort Constants as Functions of the Age of OB Stars}
\begin{center}\begin{tabular}{|c|c|r|r|r|r|c|}\hline
     Age, & $n_\star$ &       A ~~~~ &      B ~~~~  &     C ~~~~   &     K ~~~~   & Source  \\
 million years &           &    km/s/kpc  &   km/s/kpc   &  km/s/kpc    &  km/s/kpc    &                      \\\hline
   $<30$  & 275 & $-8.5\pm2.7$ & $-24.5\pm2.7$ & $10.5\pm2.7$ & $ 7.4\pm2.7$ & [85] \\
   $<30$  & 144 & $-6.1\pm4.1$ & $-20.6\pm5.2$ & $ 2.9\pm3.7$ & $11.0\pm3.5$ & [86] \\
   $<30$  & 361 & $ 5.7\pm1.4$ & $-20.7\pm1.4$ & $ 5.2\pm1.4$ & $ 7.1\pm1.4$ & [30] \\\hline
   $>60$  & 445 & $15.1\pm3.6$ & $-11.8\pm3.6$ & $-9.2\pm3.6$ & $-2.5\pm3.6$ & [86] \\
 not members & 291 & $13.7\pm1.0$ & $-13.6\pm0.8$ & $ 0.8\pm1.1$ & $-1.1\pm0.8$ & [86] \\
   $>60$  & 932 & $11.8\pm1.5$ & $-11.0\pm1.4$ & $-0.9\pm1.5$ & $-3.5\pm1.7$ & [30] \\\hline
 0.6--2~kpc &449& $13.0\pm0.7$ & $-12.1\pm0.7$ & $ 0.5\pm0.8$ & $-2.9\pm0.6$ & [30] \\\hline
 \end{tabular}\end{center}
 \end{table} }
%%%%%%%%%%%%%%%%%%%%%%%%%%%%%%%%%%%%%%%%%%%%%%%%%%%%%%%%%%%%%%%%%%%%%%%%%%%%%%%%%%%%%%

Table 1 lists the parameters of the linear Ogorodnikov-Milne model
and the Oort constants $A,B,C,$ and $K$ [85,86,30] for members
(upper part of the table) and non-members (middle of the table) of
the Gould belt. Nearby stars in a neighborhood with a radius of
$\approx$600 pc were considered. In the lowest section of the
table the Oort constants $A,B,C,$ and $K$ characterize the
differential rotation of the Galaxy. The second column gives the
number of stars, $n_\star$. It can be seen from the table that the
data from various authors show consistently that the Oort
parameters for the stars in the Gould belt differ significantly
from the parameters for the galactic rotation.

Based on a sample of 49 nearby diffuse stellar clusters with an
average age of 32 million years, Bobylev [88] has shown that the
Gould belt participates in several motions. First, besides
involvement in the common rotation of the Galaxy, the entire
complex as a whole moves relative to the local standard of rest at
a velocity of 10.7$\pm$0.7 km/s in the $l=274^\circ\pm4^\circ$,
$b=-1^\circ\pm3^\circ$ direction. Second, there is a residual
rotation and expansion of the system. The parameters of the
kinematic center were taken to be $l_0=128^\circ$ and
$R_0=150$~pc. The residual velocities reach a maximum of
$-4.3\pm1.9$~km/s for the rotation and 4.1$\pm$1.4~km/s for the
expansion with a distance from the kinematic center of
$\approx$300~pc.

%%%%%%%%%%%%%%%%%%%%%%%%%%%%%%%%%%%%%%%% f.6:
 \begin{figure}[t]
 {\begin{center}
 \includegraphics[width=120mm]{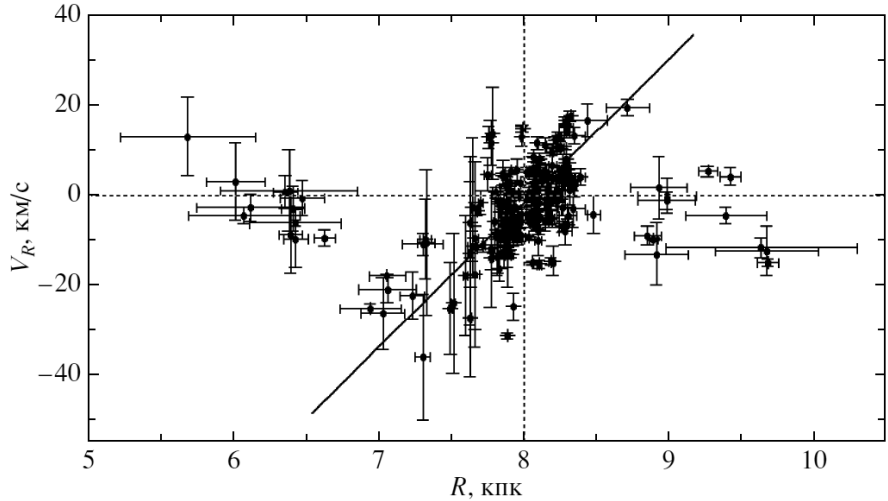}
 \caption{
 Galactocentric radial velocities for a sample of 220 stars of spectral
classes O--B2.5 as a function of the galactocentric distance R
according to the data of Ref.~[89]. The sun’s position is
indicated by a vertical dashed line.}
 \label{fig6}
 \end{center} }
 \end{figure}
%%%%%%%%%%%%%%%%%%%%%%%%%%%%%%%%%%%%%%%%%%%%%%%

We note that accounting for the influence of the rotation of the
Galaxy plays an important role in analyzing the kinematics of the
Gould belt. However, the parameters of the spiral pattern, such as
the number of arms, twist angle, rotational velocity of the
pattern, and phase of the sun in the spiral wave are poorly known.
Bobylev and Bajkova have studied [89] a sample of 220 stars, some
of which were mostly distant stars of spectral classes O-B2.5
while the others (belonging to the Gould belt) were massive
HIPPARCOS B-stars with parallax errors of no more than 10\%.
Figure 6 shows the galactocentric radial velocities of these stars
as functions of their galactocentric distance R. A wave with a
length of about 3 kpc and an azimuth of about 10~km/s shows up
clearly. It is related to the influence of a galactic spiral
density wave in the velocities of the stars. It was shown that the
perturbation in the radial velocities has a similar phase for both
distant and nearby stars. A line corresponding to a velocity
gradient of $dV_R/dR=40$~km/s/kpc is drawn in the figure. This
gradient is part of the classical kinematic K-effect, which we
mentioned at the very beginning of this article:
  $
  K=0.5\biggl[
   {\strut \displaystyle \partial\displaystyle V_R\over\displaystyle\partial\displaystyle R}
  +{\strut \displaystyle 1\over\displaystyle R}{\strut\displaystyle\partial\displaystyle V_\theta\over\displaystyle\partial\displaystyle \theta}
  +{\strut\displaystyle  V_R\over\displaystyle R}\biggr].
  $
This straight line shows that the ``observed'' expansion of the
Local system of stars (the Gould belt, in particular) is
essentially a manifestation of a local perturbation caused by the
spiral density wave. The parameters of the velocity of the
perturbation from the spiral wave in the residual rotation
velocities of the stars are different for the nearby (Gould belt)
and distant stars. This is especially noticeable from the phase of
the sun in the spiral wave. This indicates that the Gould belt may
have an intrinsic residual rotation that is unrelated to the
influence of the spiral density wave.

 \subsection{Age}
 Various methods have been used to estimate the age
of the Gould belt. Frogel and Stothers [21] have written in
interesting review of these methods. Here we supplement their
approach by adding some current results.

Estimates of the age of the Gould belt based on the measured
expansion coefficient for stars and for gas clouds
[21,60,85,87,90--94] yield values in the range of 30--70 million
years. (A wider range of 30--220 million years is given in the
above review [21].)

Estimates of the age of the Gould belt based on the motion of the
vertex [9] lie in the range of 20--70 million years. Estimates of
the age of individual stars in the Gould belt using, for example,
Str\"{o}mgren photometry [30], show that stars with ages
$<$90~million years belong to this structure. The ages of diffuse
stellar clusters and OB associations have been estimated by
comparison with isochrones which show that the diffuse stellar
clusters with ages $<$80 million years belong to the Gould belt
[36]; the average ages of the individual OB associations do not
exceed 50 million years [42].

All of the above results are in good agreement among themselves.
At present, it is usually assumed that the average age of the
Gould belt is $\approx$60 million years.

A number of authors have found an intrinsic rotation of the Gould
belt [48,95,61,94,88]. Estimates of the age of the Gould belt
obtained using the angular velocity of its rotation yield values
in the range of 50-500 million years and differ most strongly from
the estimates based on the other methods. We note that a value of
$\approx$80 million years, close to the lowest value in this
range, based on the angular rotation velocity of the Local system,
$-1.63"/$(100 yr)$=-77$~km/s/kpc, was obtained by Shatsova [48].
An analysis of modern data does not confirm such a large value for
this quantity. Modern determinations of the intrinsic angular
rotation velocity of the Gould belt give a value between $-25$ and
$-20$~km/s/kpc [61,88] and in this case the estimated rotation
period is 200$\div$300 million years. This paradox has not yet
been resolved.

 \subsection{Mass}
Table 2 lists the estimated mass of the Gould belt and its
individual components obtained by various authors based on
different data and methods.

Original estimates of the mass of neutral hydrogen and dust were
obtained by Davis [58] using their concentration as derived from
an extensive program of observations of radio-frequency hydrogen
lines. They obtained a preliminary estimate for the mass of all
the stars in the different spectral classes belonging to the Gould
belt.

Lindblad estimated [59] the mass of neutral hydrogen on the basis
of an analysis of original 21 cm radio observations. The estimate
of the mass of neutral hydrogen in the paper by Olano [96] was
based on assuming a supernova explosion, followed by expansion and
slowing down of the resulting shell. The mass of the Gould belt
has been estimated by others [32,12,93] beginning with similar
assumptions. They have pointed out that the difference between
these estimates and that of Olano [96] mainly arises from the use
of different values for the density of stars in the vicinity of
the sun.

%%%%%%%%%%%%%%%%%%%%%%%%%%%%%%%%%%%%%%%%%%%%%%%%%%%%%%%%%%%%%%%%%%%%%%%%%%%%%
{\begin{table}[t]
\small                                               %% t2
\caption[]{\small
 Estimates of the Mass of the Gould Belt and its Individual Components}
\begin{center}\begin{tabular}{|r|c|r|r|c|r|r|}\hline
  Component &          Age, &      Radius,  &         Mass, ~   &  Source  \\
            & million years &      pc~~~    &     $M_\odot$ ~~~ &  \\\hline
      Stars &      --- &  $\approx500$ &       $10^4-10^5$ &  [58] \\
            &  $45-90$ &         $500$ &     $5\times10^5$ &  [32] \\\hline
      HI    &      --- &  $\approx500$ &   $2.6\times10^5$ &  [58] \\
            &     $60$ &  $\approx600$ &     $1\times10^6$ &  [59] \\
            &     $30$ &  $\approx300$ &   $1.2\times10^6$ &  [96] \\
            &     $16$ &  $\approx300$ & ~~$3.3\times10^5$ &  [12] \\
            &     $26$ &  $\approx300$ &   $2.4\times10^5$ &  [93] \\\hline
      H$_2$ & $\leq60$ &  $\approx300$ &     $4\times10^5$ &  [32] \\\hline
     Dust   &      --- &  $\approx500$ &     $6\times10^3$ &  [58] \\\hline
Mass in center & $20-40$ &  $<500$ & $\approx1\times10^6$ &  [61] \\
               &    $32$ &  $<500$ &      $1.5\times10^6$ &  [88] \\\hline
 \end{tabular}\end{center}
 \end{table} }
%%%%%%%%%%%%%%%%%%%%%%%%%%%%%%%%%%%%%%%%%%%%%%%%%%%%%%%%%%%%%%%%%%%%%%%%%%%%%%%%%%%%%%

To estimate the overall mass of molecular H2 clouds belonging to
the Gould belt, Taylor, et al. [32], have reanalyzed the data of
Lynds [97] taking into account the redistribution of matter that
is typical of the Gould belt. The total mass of the stars in the
Gould belt was estimated quite carefully, using an initial mass
function derived from the data of various authors.

Lindblad [61] and Bobylev [88] have made virial estimates—Lindblad
by analyzing the kinematics of the youngest fraction of the
HIPPARCOS OB stars and Bobylev by analyzing the rotation curve for
49 diffuse stellar clusters and associations belonging to the
Gould belt. This approach is based on the assumption that all of
the mass is concentrated in the center, while the motion of the
stars obeys the Kepler law. This method yields an estimate for all
the gravitational mass within a specified volume of space.

Table 2 shows that the major contribution to the mass of the Gould
belt is from neutral hydrogen. Thus, the Gould belt is properly
referred to as a gas-stellar complex.

 \section{Formation scenarios}
A number of scenarios have been proposed for the formation of the
Gould belt. According to one, it was formed as the result of a
supernova explosion. According to a second scenario, it was formed
as the result of the collision of high-velocity clouds of neutral
hydrogen with the Galactic disk. According to a third, the
formation of the Gould belt is a stage in the kinematic evolution
of the Orion arm.

As noted by P\"{o}ppel [26], on the whole the star-formation
process in the sun’s surroundings could have been provoked by the
passage of the Carina-Sagittarius arm through this region. A
process of spontaneous star formation and its propagation could
also have played a role.

 \subsection{Supernova explosion}
This approach is based on Blaauw's proposal [98] that the Gould
belt could have formed as the result of the expansion of extremely
hot gas from a very small spatial volume, i.e., an explosion. An
explosion of this sort would produce an expanding shell. As a
source of stellar wind or an explosion, Blaauw [99,100] examined
the stars in the Cas-Tau OB association. At present, this
association is spread out over a substantial space, but the
cluster a Per lies at its center (Fig. 2). As a whole, the
distribution of the nearest OB associations belonging to the Gould
belt [100] does not entirely match the predictions of this model,
so that Blaauw [100] concludes that the model is not complete.

Nevertheless, based on the model of a supernova expression, a
number of important results have been obtained by Olano [96],
Moreno, et al. [12], P\"{o}ppel and Marronetti [101], and Perrot
and Grenier [93].

Olano [96] examined a gas dynamic model for the formation of the
Gould belt with a substantial initial expansion velocity
($\approx$20~km/s) of an initial hydrogen cloud. Because the gas
slows down owing to the drag of the surrounding medium, the zero
velocity limit outlines the outer boundary of the Gould belt. An
ellipse with $\approx$20 pc semiaxies centered at $l_0=131^\circ$
and $R_0=166$ pc was found (Fig. 2). The other parameters found in
Olano's paper have already been mentioned above.

Lindblad [61] has proposed a model of intrinsic differential
rotation and expansion of the Gould belt which was considered as a
gravitationally-coupled system with an angular velocity of
$\omega_0=-24$~km/s/kpc coincident with the direction of the
galactic rotation along with expansion of the system with an
angular velocity coefficient $\rho_0=­20$~km/s/kpc for the found
center parameters $l_0=127^\circ$ and $R_0=166$~pc. This model
takes into account an inclination of the disk to the galactic
plane of $20^\circ$ and was constructed using results from an
analysis of the HIPPARCOS OB stars obtained in Refs. 102 and 87.
As opposed to the Olano model [96], the Lindblad model [61]
explains the flat shape of the Gould belt in terms of its having a
substantial angular momentum. Bobylev [94,88] extended the
Lindblad approach to the nonlinear case with an exact calculation
of the distance from the kinematic center of the system to stars
(using the measured parallaxes of the stars); similar values of
the kinematic parameters were obtained.

Perrot and Grenier [93] used radial velocities and distances of
the molecular clouds based on a three-dimensional model for the
evolution of the expanding shell and obtained results that are, on
the whole, in satisfactory agreement with the Olano model [96].
Somewhat different parameters of the ellipse,
$\approx373\times233\times30$~pc with a center at $l_0=180^\circ$
and $R_0=104$~pc were obtained, as well as a substantially lower
estimate for the mass of hydrogen (Table 1). They reached the
interesting conclusion that the current geometric and kinematic
characteristics of the Gould belt are essentially independent of
the initial rotation direction. Confirming Blaauw's opinion [100],
they note the following contradiction. The explosion model assumed
that the older OB associations were formed as a result of the
interaction of a faster shock wave than the younger ones. This
means that the older associations must lie further from the center
of the explosion and move at higher speeds than the younger ones.
However, the expected velocity gradient and distances are not
observed in the Scorpio-Centaurus (US, UCL, and LCC) and Orion
(Ori 1a, Ori 1b, and Ori 1c) associations, for which reliable
estimates of the age of the individual groups are available.

Palou\v{s} [103] has modelled three cases in the framework of the
explosion scenario: (a) free expansion from a point center, (b)
the development of a shell similar to that observed around the OB
associations, and (c) the development of a shell that appears as
the result of the explosion of a hypernova (a powerful explosion
resulting, for example, from the merger of two neutron stars).
This author concludes that the observed characteristics of the
Gould belt, specifically the Oort constants and the shape of the
belt, cannot be explained in terms of these models. In case (a)
the Oort constant B remains equal to 0 during the evolution of the
shell; in case (b) the shell does not break up into fragments, in
conflict with observations; and, in case (c) the development of
the shell leads to a very prolate figure which agrees very poorly
with the observed shape of the Gould belt.

Based on a comparison of observational data on stars with models
of the expansion of the Gould belt in Refs. 91 and 104, the
authors conclude that the expansion is more likely from a line,
than a point center.

In the explosion model the following points are not clear: why the
complex of molecular clouds in Taurus lies inside the expanding
Lindblad ring (see Fig. 4) and what kind of role a magnetic field
may play in bubble formation [26].

Nevertheless, digressing from the physical causes of the initial
interaction with the parent cloud, it is possible reproduce the
major features of the further evolution of the Gould belt. For
example, based on a numerical simulation of the dynamic evolution
of the Gould belt in a suitable galactic potential, Vasil'kova
[105] has shown that an initial spherical distribution of the
model particles becomes ellipsoidal with time (in accord with
Blaauw’s earlier results) and that collective oscillations of the
particles along the z axis occur which are typical of the Gould
belt.

 \subsection{High-velocity clouds}
Surveys of the distribution and motion of hydrogen show that the
velocities of almost all the high-latitude hydrogen clouds are
such that they converge toward the Galactic plane, and in a number
of cases their velocities reach 200 km/s or more. Although the
distances to these clouds are poorly known, it is assumed that
they are associated with the Magellan flow [106,107].

L\'epin and Duver [108] have proposed that a number of the large
complexes of molecular clouds which lie sufficiently far from the
galactic plane could be formed as a result of a collision of
high-velocity clouds with the Galactic disk. A simple
two-dimensional magnetohydrodynamic model was used to explain the
appearance of such observed clusters of molecular clouds as Ori,
Cha, r Oph, and Tau-Aur. Each cluster was treated separately, so
that a structure such as the Gould belt could appear randomly in
this model.

Comer\'on and Torra [109,104] have examined a more complicated
model of oblique incidence of a high-velocity cloud on the
galactic plane. They showed that a structure similar to the
Lindblad ring ultimately forms, but with considerably larger
dimensions. As P\"{o}ppel [26] noted, the study of these processes
is of great interest for understanding the origin of the Gould
belt, although the models proposed thus far encounter a number of
problems in explaining some of the properties of the interstellar
medium.

It is interesting to note the model of Bekki [110], which is
similar to the model of Comeron and Torra, but involves a cloud of
dark matter instead of a high-velocity hydrogen cloud. Bekki's
numerical simulations show that the Gould belt could have been
formed about 30 million years ago from an initial gas cloud with a
mass of about $10^6 M_\odot$ after a collision with a clump of
dark matter with a mass of $3\times10^7 M_\odot$. In the
calculations, the clump of dark matter moves from the southern
into the northern hemisphere at an angle of about 30$^\circ$ to
the galactic plane. Its dynamic influence is such that a star
formation process starts in an initially symmetric gas cloud (the
parent of the Gould belt), the cloud gradually elongates into an
ellipse, and finally acquires the dimensions and inclination to
the galactic plane characteristic of the Gould belt.

 \subsection{Evolution of the Orion arm}
In a paper by Olano [44] the Local system is identified with the
Local arm (Orion arm) and the evolution of this structure with a
mass of $2\times10^7 M_\odot$ is modelled over the last 100
million years. In this approach the gas moves at a high initial
velocity ($\approx$50 km/s) and the Local system is formed from
it. It is assumed that this velocity could be achieved by an
interaction with the Carina-Sagittarius arm. A collision of a gas
cloud with a spiral density wave would lead to its breakup. In
terms of this model, such clusters as the Hyades, Pleiades, and
Coma Berenices, and the Sirius cluster are treated as fragments of
a once unified complex and the compression of the central regions
of a parent cloud led to the formation of the Gould belt. The
construction of orbits with different structures in the Orion arm
led to an interesting result: it turned out that the gravitation
of the Local system has a significant effect on the sun's motion.

 \section{Conclusion}
At present there is a large scale earthbound survey of the sky
(RAVE) under way for the purpose of determining the radial
velocities of hundreds of thousands of stars, as well as a space
experiment (GAIA) which will yield an enormous base of
high-precision data on the distances and proper motions of
millions of stars with microsecond accuracy, their radial
velocities with accuracies of a fraction of a km/s, and their
photometry. VLBI observations of galactic masers are continuing
for the purpose of determining their proper motions and
trigonometric parallaxes with high accuracy. These projects are
primarily intended for studying the structure and kinematics of
the Galaxy, since they will significantly expand the possibilities
for studying the three-dimensional motions of stars lying at
distances of up to 10 kpc from the sun.

As for studies of the Local system, the availability of high
accuracy data should primarily enable reliable separation of
objects in the Gould belt and Orion arm from the surrounding
background in terms of an entire series of parameters – age,
distribution, distance, kinematics. As a whole, this will aid in
solving some of the following problems:

 {\begin{itemize}
\item clarifying the reasons for the initial interaction (or set
of these) which led to the compression of a parent gas cloud from
which the Orion arm and the Gould belt were formed;

\item constructing an adequate dynamic model for the evolution of
the Orion arm and the Gould belt; and,

\item a detailed reconstruction of the history of star formation
within the Local system.
\end{itemize}}

A theory of the origin of the Gould belt should explain the
following: its inclination; the distribution of the OB
associations systems of molecular clouds surrounding it; the
features of its three-dimensional velocity distribution; and, aid
in clarifying its gravitational coupling.

The author thanks V. P. Grinin for initiating the writing of this
review and V. V. Vityazev for attentive reading of the manuscript
and comments that helped improve the paper. This work was
supported by program P--21 of the Russian Academy of Sciences,
``Nonstationary phenomena in objects in the universe.''

 \bigskip{REFERENCES}
 \bigskip
 {\small

 1. J.F.W. Herschel, Results of astronomical observations made
during the years 1834, 5, 6, 7, 8 at the Cape of Good Hope, Smith,
Elder and Co., London (1847).

 2. F.G.W. Struve, \'Etudes d’astronomie stellaire. Sur la voie
 lact\'ee et sur la distance des etoiles fixes, St. Petersburg (1847).

 3. B.A. Gould, Proc. of the American Assoc. for Advanced Sci. Part I, 115 (1874).

 4. B.A. Gould, Uranometria Argentina, P.E. Coni, Buenos Aires (1879).

 5. H. Shapley, Astrophys. J. 49, 311 (1919).

 6. J.C. Kapteyn and E.B. Frost, Astrophys. J. 32, 83 (1910).

 7. W.W. Campbell, Lick Obs. Bull. 6, 101 (1911).

 8. E.A. Milne, Mon. Notic. Roy. Astron. Soc. 95, 560 (1935).

 9. K.F. Ogorodnikov, Dynamics of Stellar Systems. Oxford: Pergamon press (1965).

 10. G. Str\"{o}mberg, Astrophys. J. 59, 228 (1924).

 11. P.P. Parenago, Astron. zh. 27, 150 (1950).

 12. E. Moreno, E.J. Alfaro, and J. Franco, Astrophys. J. 522, 276 (1999).

 13. B. Lindblad, Mon. Notic. Roy. Astron. Soc. 87, 553 (1927).

 14. B. Lindblad, Mon. Notic. Roy. Astron. Soc. 90, 503 (1930).

 15. J.H. Oort, Bull. Astron. Inst. Netherlands 3, 120, 275 (1927).

 16. J.H. Oort, Bull. Astron. Inst. Netherlands 4, 132, 79 (1927).

 17. R.J. Trumpler, Lick Obs. Bull. 14, 420 (1930).

 18. V.A. Ambartsumyan, The Evolution of Stars and Astrophysics
[in Russian], Izd. AN Arm. SSR, Erevan (1947).

 19. V.A. Ambartsumyan, Astron. zh. 26, 3 (1949).

 20. B.J. Bok, The Distribution of the Stars in Space, U. Chicago Press, Chicago (1937).

 21. J.A. Frogel and R. Stothers, Astron. J. 82, 890 (1977).

 22. Yu.N. Efremov, Star-formation Centers in Galaxies [in Russian], Nauka, Moscow (1989).

 23. W.G.L. P\"oppel, Fundamentals of Cosmic Physics, 18, 1--271 (1997).

 24. Yu.N. Efremov, Astron. Astrophys. Trans. 15, 3 (1998).

 25. Yu.N. Efremov and B.G. Elmegreen, Mon. Notic. Roy. Astron. Soc. 299, 588 (1998).

 26. W.G.L. P\"oppel, ASP Conf. Ser. 243, 667 (2001).

 27. M.F. Skrutskie, R.M. Cutri, R. Stiening, et al., Astron. J. 131, 1163 (2006).

 28. The HIPPARCOS and Tycho Catalogues, ESA SP--1200 (1997).

 29. E. Hog, C. Fabricius, V.V. Makarov, et al., Astron. Astrophys. 355, L27 (2000).

 30. J. Torra, D. Fern\'andez, and F. Figueras, Astron. Astrophys. 359, 82 (2000).

 31. E. Chereul, M. Creze, and O. Bienayme, Astron. Astrophys. 340, 384 (1998).

 32. D.K. Taylor, R.L. Dickman, and N.Z. Scoville, Astrophys. J.315, 104 (1987).

 33. R. Wichmann, J.H.M.M. Schmitt, and S. Hubrig, Astron. Astrophys. 399, 983 (2003).

 34. E.E. Mamajek, M. Meyer, and J. Liebert, Astron. J. 124, 1670 (2002).

 35. J. Hern\'andes, N. Calvet, L. Hartmann, et al., Astron. J. 129, 856 (2005).

 36. A.E. Piskunov, N.V. Kharchenko, S. Roser, et al., Astron. Astrophys. 445, 545 (2006).

 37. J.H.J. de Bruijne, Mon. Notic. Roy. Astron. Soc. 310, 585 (1999).

 38. S. Madsen, D. Dravins, and L. Lindegren, Astron. Astrophys. 381, 446 (2002).

 39. A. Blaauw, Bull. Astron. Inst. of Netherlands 11, 414 (1952).

 40. A. Blaauw, Ann. Rev. Astron. Astrophys. 2, 213 (1964).

 41. M.J. Sartori, J.R.D. L\'epine, and W.S. Dias, Astron. Astrophys. 404, 913 (2003).

 42. P.T. de Zeeuw, R. Hoogerwerf, J.H.J. de Bruijne, et al., Astron. J. 117, 354 (1999).

 43. A.G.A. Brown, E.J. de Geus, and P.T. de Zeeuw, Astron. Astrophys. 289, 101 (1994).

 44. C.A. Olano, Astron. Astrophys. 121, 295 (2001).

 45. Yu.N. Efremov and T.G. Sitnik, Pis’ma v Astron. zh. 14, 817 (1988).

 46. H. Mineur, Mon. Notic. Roy. Astron. Soc. 90, 516 (1930).

 47. H. Mineur, Mon. Notic. Roy. Astron. Soc. 90, 789 (1930).

 48. R.B. Shatsova, Uchenye zapiski LGU 136, No. 22, 113 (1950).

 49. A.S. Tsvetkov, Astron. Astrophys. Transactions 8, 145 (1995).

 50. A.S. Tsvetkov, Astron. Astrophys. Transactions 9, 1 (1995).

 51. A.S. Tsvetkov, in: J. Vondr\'ak and N. Capitaine, eds., Proc.
of Conf. ``Jornees Systemes de reference spatio temporeles 1997''
Prague, 22--24 September, p. 171 (1997).

 52. O.J. Eggen, Publ. Astron. Soc. Pacif. 87, 37 (1975).

 53. O.J. Eggen, Publ. Astron. Soc. Pacif. 89, 187 (1977).

 54. D. Montes, J. L\'opez-Santiago, M.C. G\'alvez, et al., MNRAS 328, 45 (2001).

 55. J. Lopez-Santiago, D. Montes, I. Crespo-Chac\'on, et al., Astrophys. J. 643, 1160 (2006).

 56. K.A. Barkhatova, L.P. Osipkov, and S.A. Kutuzov, Astron. zh. 66, 1154 (1989).

 57. A.E. Lilley, Astrophys. J. 121, 559 (1955).

 58. R.D. Davies, Mon. Not. Roy. Astron. Soc. 120, 35 (1960).

 59. P.O. Lindblad, Bull. Astron. Inst. Netherland 19, 34 (1967).

 60. P.O. Lindblad, K. Grape, A. Sandqvist, et al., Astron. Astrophys. 24, 309 (1973).

 61. P.O. Lindblad, Astron. Astrophys. 363, 154 (2000).

 62. T.M. Dame, D. Hartmann, and P. Thaddeus, Astrophys. J. 547, 792 (2001).

 63. H.C. Bhatt, Astron. Astrophys. 362, 715 (2000).

 64. A. Porras, M. Christopher, L. Allen, et al., Astron. J. 126, 1916 (2003).

 65. M.J. Reid, K.M. Menten, X.W. Zheng, et al., Astrophys. J. 700, 137 (2009).

 66. K.L.J. Rygl, A. Brunthaler, M.J. Reid, et al., Astron. Astrophys. 511, A2 (2010).

 67. A. Brunthaler, M.J. Reid, K.M. Menten, et al., AN 332, 461 (2011).

 68. T. Hirota, T. Bushimata, Y.K. Choi, et al., PASJ 59, 897 (2007).

 69. R.M. Torres, L. Loinard, A.J. Mioduszewski, et al., Astrophys. J. 671, 1813 (2007).

 70. L. Loinard, R.M. Torres, A.J. Mioduszewski, et al., Astrophys. J. Lett. 675, L29 (2008).

 71. M.J. Reid, K.M. Menten, A. Brunthaler, et al., Astrophys. J. 783, 130 (2014).

 72. Y. Xu, J.J. Li, M.J. Reid, et al., Astrophys. J. 769, 15 (2013).

 73. M.P. Fitzgerald, Astron. J. 73, 983 (1968).

 74. D.M. Sfeir, R. Lallement, F. Grifo, et al., Astron. Astrophys. 346, 785 (1999).

 75. T. Preibish and H. Zinnecker, Astron. J. 117, 2381 (1999).

 76. T.W. Berghofer and D. Breitschwerdt, Astron. Astrophys. 390, 299 (2002).

 77. S.B. Popov, M. Colpi, M.E. Prokhorov, et al., Astron. Astrophys. 406, 111 (2003).

 78. C. Motch, A.M. Pires, F. Haberl, et al., Astrophys. Space Sci. 308, 217 (2006).

 79. P.C. Frisch, Space Sci. Rev. 72, 499 (1995).

 80. J. Maiz-Apell\'{a}niz, Astrophys. J. 560, L 83 (2001).

 81. D. Breitschwerdt and M.A. de Avillez, Astron. Astrophys. 452, L1 (2006).

 82. C. Heiles, Astrophys. J. 498, 689 (1998).

 83. D. Ward-Tompson, J. Di Francesco, J. Hatchell, et al., PASP 119, 855 (2007).

 84. P.C. Frisch, Space Sci. Rev. 130, 355 (2007).

 85. T.N.G. Westin, Astron. Astrophys. Suppl. Ser. 60, 99 (1985).

 86. P.O. Lindblad, J. Palou\v{s}, K. Loden, et al., in: B.
Battrick, ed., HIPPARCOS Venice’97, ESA Publ. Div., Noordwijk, p.
507 (1997).

  87. F. Comer\'on, Astron. Astrophys. 351, 506 (1999).

  88. V.V. Bobylev, 2006, Astron. Lett., 32, 816 (2006).

  89. V.V. Bobylev and A.T. Bajkova, Astron. Lett., 39, 532 (2013).

  90. S.V.M. Clube, Mon. Notic. Roy. Astron. Soc. 137, 189 (1967).

  91. J.R. Lesh, Astrophys. J. Suppl. Ser. 17, 371 (1968).

  92. A. Tsioumis and W. Fricke, Astron. Astrophys. 75, 1 (1979).

  93. C.A. Perrot and I. A. Grenier, Astron. Astrophys. 404, 519 (2003).

  94. V.V. Bobylev, Astron. Lett., 30, 784 (2004).

 95. K.F. Ogorodnikov, Trudy astron. observatorii LGU 15, 1 (1950).

 96. C.A. Olano, Astron. Astrophys. 112 (195 (1982).

 97. B.T. Lynds, Astrophys. J. Suppl. Ser. 7, 1 (1962).

 98. A. Blaauw, Koninkl. Ned. Akad. Wetenschap. 74, 4 (1965).

 99. A. Blaauw, Astrophys. J. 123, 408 (1956).

 100. A. Blaauw, Physics of Star Formation and Early Stellar Evolution, N. Kalifas
and Ch. Lada, eds., Kluwer Acad. Publ., Dordrecht (1991).

 101. W.G.L. P\"{o}ppel and P. Marronetti, Astron. Astrophys. 358, 299 (2000).

 102. J. Torra, A.E. G\'omez, F. Figueras, et al., in: B. Battrick,
ed., HIPPARCOS Venice’97, ESA Publ. Div. Noordwijk, p. 513 (1997).

 103. J. Palou\v{s}, Astrophys. Space Sci. 276, 359 (2001).

 104. F. Comer\'on and J. Torra, Astron. Astrophys. 281, 35 (1994).

 105. O.O. Vasil’kova, Pis’ma v Astron. zh. 40, 63 (2014).

 106. C.A. Olano, Astron. Astrophys. 423, 895 (2004).

 107. C.A. Olano, Astron. Astrophys. 485, 457 (2008).

 108. J.R.D. L\'epin and G. Duvert, Astron. Astrophys. 286, 60 (1994).

 109. F. C\'omeron and J. Torra, Astron. Astrophys. 261, 94 (1992).

 110. K. Bekki, Mon. Notic. Roy. Astron. Soc. 398, L36 (2009).

 }

\end{document}